\documentclass[a4paper,12pt]{article}
\usepackage{amsthm}
\usepackage{amssymb}
\usepackage{amsmath}
\usepackage{amsfonts}
\usepackage{color}
\usepackage{graphicx}
\usepackage[square, comma, sort&compress, numbers]{natbib}

\newtheorem{lemma}{Lemma}[section]

\newtheorem{proposition}{Proposition}[section]

\newtheorem{definition}{Definition}[section]

\def\b1{\mbox{\boldmath $1$}}

\newenvironment{demo*}{\vspace{3mm}\noindent{\bf Proof.}}{\hfill $\Box$ \vspace{3mm}}

\begin{document}
\title{\bf \Large {A new multivariate dependence measure based on comonotonicity}}
{\color{red}{\author{\normalsize{Ying Zhang}\\
{\normalsize\it  School of Mathematical Sciences, Qufu Normal University}\\
\noindent{\normalsize\it Shandong 273165, China}\\
e-mail:  zhangying0513@gmail.com\\
\and
\normalsize{Chuancun Yin}
\thanks{Corresponding author.}\\
{\normalsize\it  School of Statistics,  Qufu Normal University}\\
\noindent{\normalsize\it Shandong 273165, China}\\
e-mail:  ccyin@mail.qfnu.edu.cn}}}
\maketitle
\vskip0.01cm
\noindent{\large {\bf Abstract}}  { {In this paper we introduce a new multivariate dependence measure based on comonotonicity by means of product moment which motivated by the recent papers of Koch and Schepper (ASTIN Bulletin 41 (2011) 191-213) and Dhaene et al. (Journal of Computational and Applied Mathematics 263 (2014) 78-87). Some differences and relations between the new dependence measure and other multivariate measures are analyzed.
We also give several characteristics of this measure and estimations based on the definitions and its property are presented.}}

\medskip



\numberwithin{equation}{section}
\section{Introduction}\label{intro}

Early in 1966, a pioneering paper by E. L. Lehmann \cite{LEL} which gave numerous useful results in both statistical theory and application started the study of concepts of dependence for random vectors. Applications of these concepts in actuarial science have recently received increased interest. Scholars have proposed many concepts to formalize the notation of dependence existing between risks. Early sources include Lehmann \cite{LEL}, Esary, Proschan and Walkup \cite{EPW}, Esary and Proschan \cite{EP} and Kimeldorf and Sampson \cite{KS}. We are grateful that Mari and Kotz \cite{MK} pointed out the concept of correlation introduced by Galton in 1885 dominated statistics until the 1970s serving as practically the only measure of dependence. This often resulted in somewhat misleading conclusions, for reasons becomes clear in the later literatures.\\
\indent The following problem which can't be avoided is how to measure a dependent random vector in practice. A risk measure is defined as a mapping from a set of random variables or vectors representing the risks at hand to the real numbers. Scarsini \cite{Scarsini} defined certain desirable properties for a measure of association between two random variables (see also Definition 5.1.7 and Theorem 5.1.8 in Nelsen \cite{Nelsen}) and coined the name concordance measures for those satisfying these conditions. There are a lots of ways to discuss and to measure dependence. First and foremost is Pearson's correlation coefficient \cite{Denuit2005} which captures the linear dependence between couples of random variables, but which is not invariant under monotone transformations of coordinate axes. Then, Kendall's tau and Spearman's rho are proposed to measure a form of dependence known as concordance, which are scale-invariant. The rank correlation coefficients fulfill all the desirable properties provided $X_1$ and $X_2$ are continuous. After that, Kaas et al. \cite{KDG} and Dhaene et al. \cite{DDGa} defined bounds in convex order sense for the multivariate dependence measure, calculated by means of comonotonic vectors. Recently Koch and Schepper \cite{KD} proposed a comonotonicity coefficient for arbitrary $m$-dimensional vectors to measure the degree of comonotonicity, they also explained why in the case of cash flows the convex bounds reveal such effective approximations in Kaas et al. \cite{KDG} and Dhaene et al. \cite{DDGa}.\\
\indent The impetus for writing this paper came while reading Dhaene et al. \cite{2014}, which defined a multivariate dependence measure for aggregating risks as follows.
  \begin{eqnarray*}
      \rho_C(\textbf{X})=\frac{\text{Var}{(S)}-\text{Var}{(S^\bot)}}{\text{Var}{(S^C)}-\text{Var}{(S^\bot)}}=\frac{\sum_{i=1}^m\sum_{j<i}\text{Cov}{(X_i,X_j)}}{\sum_{i=1}^m\sum_{j<i}\text{Cov}{(X_i^C,X_j^C)}}
           \end{eqnarray*}
     provided the covariances exist.
     In this definition, without any doubt, the calculation is largely simplified. However, the dependence between each pair of components is considered while ignoring the dependence among all the components. Especially in the case that all pairs of components are uncorrelated, the dependence measures are always 0. So in this paper we introduce a new multivariate dependence measure $\rho$ that takes all aspects into account. Then, we also find the new dependence measure is surprisingly similar to Koch and Schepper's comonotonicity coefficient $\kappa$ in \cite{KD} when it comes to non-negative random vectors. However, the new one focus on the tail distribution function rather than on the distribution function. Of course, the new measure is also different from other multivariate dependence measures in e.g. \cite{KD}-\cite{SS} when the dimension $m\geq3$. When $m=2$ we can relate our new dependence measure to the classical ones.\\
    \indent The paper is organised as follows. In Section 2 we recall some definitions and notions on distributions and copulas, and we repeat the most common dependence measures. Section 3 is the most important part, we introduce our new dependence measure and  discuss its major properties. Special attentions are given to the relationships between the new dependence measure and Koch and Schepper's comonotonicity coefficient \cite{KD} and Dhaene et al.'s dependence measure \cite{2014} in Section 4. Afterwards in Section 5 we give some comments on the estimation.

 \vskip 0.2cm
 \section{ Preliminaries}
\setcounter{equation}{0}

First of all, we briefly review the Fr\'echet space in this section, which is an essential concept when investigating the dependence structure in multivariate vectors.\\
\begin{definition}
(Fr\'{e}chet space) Let $F_1,F_2,...,F_m$ be univariate $dfs$. The Fr\'{e}chet space $\Re_m(F_1,F_2,...,F_m)$ consists of all the $m$-dimensional ($dfs$  $F_\textbf{X}$ of) random vectors $\textbf{X}$ possessing $F_1,F_2,...,F_m$ as marginal dfs, that is,
\begin{equation}
F_i(x)=P(X_i\leq x),~x\in\mathbb{R},~i=1,2,...,m.\nonumber
\end{equation}
\end{definition}
\indent Define the Fr\'{e}chet upper bound as
\begin{equation}
W_m(\textbf{x})=\min{\{F_1(x_1),F_2(x_2),...,F_m(x_m)\}},~\textbf{x}\in\mathbb{R}^m,\nonumber
\end{equation}
and the Fr\'{e}chet lower bound as
\begin{equation}
M_m(\textbf{x})=\max{\{\sum_{i=1}^mF_i(x_i)-n+1,~0}\},~ \textbf{x}\in\mathbb{R}^m.\nonumber
\end{equation}
Then the inequality
\begin{equation}
M_m(\textbf{x})\leq {F_\textbf{X}(\textbf{x})}\leq {W_m(\textbf{x})}\nonumber
\end{equation}
holds for all $\textbf{x}\in\mathbb{R}^m$ and $\textbf{X}\in \Re_m(F_1,F_2,...,F_m)$.\\\nonumber
\indent  An important concept of positive dependence in a multivariate context is positive orthant dependence.\\
\begin{definition}(POD) The random vector $\textbf{X}=(X_1,X_2,...,X_m)$ is said to be positively lower orthant dependence (PLOD) if
\begin{equation}{\label{121}}\setcounter{equation}{1}
P(X_1\leq {x_1},X_2\leq {x_2},...,X_m\leq {x_m})\geq{\prod_{i=1}^m{P(X_i\leq {x_i})}}, ~\forall(x_1,x_2,...,x_m)\in \mathbb{R}^m,
\end{equation}
and it is said to be positively upper orthant dependence (PUOD) if
\begin{equation}{\label{122}}\setcounter{equation}{2}
P(X_1> {x_1},X_2> {x_2},...,X_m> {x_m})\geq{\prod_{i=1}^m{P(X_i> {x_i})}}, ~\forall(x_1,x_2,...,x_m)\in \mathbb{R}^m.
\end{equation}
The random vector $\textbf{X}=(X_1,X_2,...,X_m)$ is said to be positively orthant dependence (POD) when it's both PLOD and PUOD.
\end{definition}
  Comonotonic vector and independent vector are two important and at the same time extreme elements of a Fr\'{e}chet space.
 \begin{definition}(Comonotonic vector and independent vector)  For every Fr\'{e}chet space \\$\Re_m(F_1,F_2,...,F_m)$, we define the independent vector $\textbf{X}^I=(X_1^I,X_2^I,...,X_m^I)$ as the vector with distribution $F_\textbf{X}^I(x_1,x_2,...,x_m)=\prod_{i=1}^mF_i(x_i)$, and the comonotonic vector  $\textbf{X}^C=(X_1^C,X_2^C,...,X_m^C)$ as the vector with distribution $F_\textbf{X}^C(x_1,x_2,...,x_m)=\min_{i=1}^mF_i(x_i)$.
 \end{definition}
 \indent In Kaas et al. \cite{2001} (Section 10.6), an alternative characterization for comonotonic random vector is discussed. A random vector $\textbf{X}$ is comontonic if and only if there exist an rv Z and non-decreasing functions $t_1,t_2,...,t_m$ such that
 \begin{equation}
 \textbf{X}=_d(t_1(Z),t_2(Z),...,t_m(Z))'.\nonumber
 \end{equation}
Note that with the inverse function defined as $F_i^{-1}(p)=\inf\{{x\in \mathbb{R}|F_i(x)\geq p\}},~p\in[0,1],$ the comonotonic vector can be easily constructed as
\begin{equation}\setcounter{equation}{3}
  \textbf{X}^C=_d\big(F_1^{-1}(U),F_2^{-1}(U),...,F_m^{-1}(U)\big),~U\thicksim U[0,1].
\end{equation}
\indent In order to construct the equivalent expression of our new dependence measure, we also need to introduce the concept of copula. A copula is a joint distribution function with uniform margins.
  One of the most fundamental results about copulas is summarized in the following lemma.
\newtheorem{thm}{Lemma}
\begin{lemma}(Sklar's Theorem)
 Every $df$ $F_\textbf{X}$ of a random vector $\textbf{X}\in \Re_m(F_1,F_2,...,F_m)$ can be represented as
 \begin{equation}{\label{1111}}\setcounter{equation}{4}
 F_{\textbf{X}}(\textbf{x})= C(F_1(x_1),F_2(x_2),...,F_m(x_m)),~ \textbf{x}\in \mathbb{R}^m,
 \end{equation}
 in terms of a copula C. If the marginals $F_1,F_2,...,F_m$ are continuous then the copula involved in $(2.4)$ is unique and explicitly given by
 \begin{equation}
 C(\textbf{u})=F_{\textbf{X}}(F_1^{-1}(u_1),F_2^{-1}(u_2),...,F_m^{-1}(u_m)),~\textbf{u}\in[0,1]^m.\nonumber
 \end{equation}
 \end{lemma}
 \indent In order to compare our new dependence measure with the common ones, we recall them here; see e.g. \cite{KD,2014,BDU,SS,DDG}.\\
 (i) Peareson correlation coefficient
 \begin{equation}
 \gamma_p(X,Y)=\frac{\text{Cov}(X,Y)}{\sqrt{\text{Var}(X)\text{Var}(Y)}};\nonumber
 \end{equation}
 (ii) Kendall's rank correlation coefficient
 \begin{equation}
 \tau(X,Y)=4\int_0^1\int_0^1C(u_1,u_2)dC(u_1,u_2)-1;\nonumber
 \end{equation}
 (iii) Spearman's rank correlation coefficient
 \begin{equation}
  \rho_s(X,Y)=12\int_0^1\int_0^1(C(u_1,u_2)-u_1u_2)du_1du_2;\nonumber
 \end{equation}
 (iv) Gini's correlation coefficient
 \begin{equation}
 G(X,Y)=2\int_0^1\int_0^1(|u_1+u_2-1|-|u_1-u_2|)dC(u_1,u_2);\nonumber
 \end{equation}
 (v) Blomqvist's correlation coefficient
 \begin{equation}
 \beta(X,Y)=4F_{\textbf{X}}(F_1^{-1}(1/2),F_2^{-1}(1/2))-1;\nonumber
 \end{equation}
(vi) Koch and Schepper's comonotonicity coefficient
\begin{equation}
\kappa(\textbf{X})=\frac{\int\cdots\int(F_{\textbf{X}}(\textbf{x})- F_{\textbf{X}}^I(\textbf{x}))d\textbf{x}}{\int\cdots\int(F_{\textbf{X}}^C(\textbf{x})- F_{\textbf{X}}^I(\textbf{x}))d\textbf{x}}\nonumber;
\end{equation}
(vii)~Dhaene et al.'s multivariate dependence measure for aggregating risks
  \begin{eqnarray*}
      \rho_C(\textbf{X})=\frac{\text{Var}(S)-\text{Var}(S^\bot)}{\text{Var}(S^C)-\text{Var}(S^\bot)}=\frac{\sum_{i=1}^m\sum_{j<i}\text{Cov}(X_i,X_j)}{\sum_{i=1}^m\sum_{j<i}\text{Cov}(X_i^C,X_j^C)}.
           \end{eqnarray*}
\section{ A new dependence measure}
\setcounter{equation}{0}

\begin{definition} The dependence measure $\rho(\textbf{X})$ of a random vector $\textbf{X}$ with non-degenerate margins is defined as
 \begin{equation}\setcounter{equation}{1}
\rho(\textbf{X})=\frac{E[\prod_{i=1}^{m}X_i]-\prod_{i=1}^{m}E[X_i]}{E[\prod_{i=1}^{m}X_i^C]-\prod_{i=1}^{m}E[X_i]}
\end{equation}
 provided the expectations exist.
 \end{definition}
\indent When it comes to non-negative random vectors, the new dependence measure has a equivalent form of the ratio of two hypervolumes.
\begin{proposition}
Let $F_\textbf{X}^C$ and $F_\textbf{X}^I$ be the joint distributions of the comonotonic and the independent vector of the Fr\'{e}chet space $\Re_m(F_1,F_2,...,F_m)$ and $\bar{F}_\textbf{X}^C$ and $\bar{F}_\textbf{X}^I$ be the joint tail distributions, respectively. For any non-negative random vector $\textbf{X}$ of $\Re_m(F_1,F_2,...,F_m)$ with joint tail distribution $\bar{F}_\textbf{X}$, the dependence measure $\rho(\textbf{X})$ can be defined as:
\begin{equation}\label{3.1}\setcounter{equation}{2}
\rho(\textbf{X})=\frac{\int\cdots\int\big(\bar{F}_\textbf{X}(\textbf{x})-\bar{F}_\textbf{X}^I(\textbf{x})\big)d\textbf{x}}
{\int\cdots\int\big(\bar{F}_\textbf{X}^C(\textbf{x})-\bar{F}_\textbf{X}^I(\textbf{x})\big)d\textbf{x}}
\end{equation}
where the integrations are performed over the whole domain of \textbf{X}.
\end{proposition}
{\bf Proof.} As we all know, the product moment of the components of an $m$-dimensional non-negative random vector $\textbf{X}$ can be written as
\begin{equation}
E[\prod_{i=1}^{m}X_i]=\int_{x_1=0}^{+\infty}\cdots\int_{x_m=0}^{+\infty}\bar{F}_{\textbf{X}}(\textbf{x})d\textbf{x}\nonumber.
\end{equation}
which concludes the proof.

 {\bf Remark 3.1.}
 In the case of continuous distribution functions, there exists a unique survival copula $\bar{C}_{\textbf{X}}$ for which
 \begin{equation}
 \bar{F}_{\textbf{X}}(x_1,...,x_m)=\bar{C}_{\textbf{X}}(\bar{F}_1(x_1),...,\bar{F}_m(x_m)).\nonumber
 \end{equation}
Inserting this into the formula for $\rho({\textbf{X}})$ in (3.2) results in the following equivalent expression for the new dependence measure:
\begin{equation}\label{33}\setcounter{equation}{3}
\rho({\textbf{X}})=\frac{\int_0^1\cdots\int_0^1(\bar{C}_{\textbf{X}}(1-u_1,...,1-u_m)-\prod_{i=1}^m(1-u_i))dF_1^{-1}(u_1)\cdots dF_m^{-1}(u_m)}
{\int_0^1\cdots\int_0^1(\bar{C}_{\textbf{X}^C}(1-u_1,...,1-u_m)-\prod_{i=1}^m(1-u_i))dF_1^{-1}(u_1)\cdots dF_m^{-1}(u_m)}.
\end{equation}
\indent Actually, the multivariate distributions $\bar{C}_{\textbf{X}}(1-u_1,...,1-u_m)$ and $C_{\textbf{X}}(u_1,...,u_m)$ are different in general.
\\\indent Our new dependence measure has several interesting properties. For instance, it satisfies the axioms of normalization, monotonicity, permutation invariance and duality in Taylor (\cite {T}) just as most dependence measures.
\begin{definition}
A random vector $\textbf{X}=(X_1,X_2,...,X_m)$ is said to be smaller than the random vector $\textbf{Y}=(Y_1,Y_2,...,Y_m)$ in the concordance order, written as $\textbf{X}\leq_c\textbf{Y}$, if both
\begin{equation}
{P(X_1\leq t_1,X_2\leq t_2,...,X_m\leq t_m)}\leq{P(Y_1\leq t_1,Y_2\leq t_2,...,Y_m\leq t_m)}\nonumber
\end{equation}
and
\begin{equation}
{P(X_1> t_1,X_2> t_2,...,X_m> t_m)}\leq{P(Y_1> t_1,Y_2> t_2,...,Y_m> t_m)}\nonumber
\end{equation}
hold for all $(t_1,t_2,...,t_m)\in\mathbb{R}^m.$\end{definition}
\begin{proposition}
For any two random vectors $\textbf{X}=(X_1,X_2,...,X_m)$ and $\textbf{Y}=(Y_1,Y_2,...,Y_m)$, $\rho$ has the following properities.
\\(i)~(normalization) If $\textbf{X}$ has comonotonic components, then $\rho(\textbf{X})=1$; if $\textbf{X}$ has independent components, then $\rho(\textbf{X})=0$.
\\(ii)~(monotonicity) If $\textbf{X}$ is smaller than $\textbf{Y}$ in the concordance order, then $\rho(\textbf{X})\leq{\rho(\textbf{Y})}$ .
\\(iii)~(permutation invariance) For any permutation $(i_1,i_2,...,i_m)$ of $(1,2,...,m)$, we have that $\rho(X_{i_1},X_{i_2},...,X_{i_m})=\rho(X_1,X_2,...,X_m)$.
\\(iv)~(duality) $\rho(-{X_1},-{X_2},...,-{X_m})=\rho(X_1,X_2,...,X_m)$.\end{proposition}
{\bf Proof}. The proofs of (i) and (iii) are straightforward.
\\(ii)
Any random vectors that are ordered in concordance order obviously have the same marginal distributions. Consequently, ${X_i}=_d{Y_i}$ and $\prod_{i=1}^{m}E[X_i]=\prod_{i=1}^{m}E[Y_i],~i=1,2,...,m$. Therefore, for any comonotonic vectors $\textbf{X}^C=(X_1^C,X_2^C,...,X_m^C)$ and $\textbf{Y}^C=(Y_1^C,Y_2^C,...,Y_m^C)$, we have ${F}_{\textbf{X}^C}(x_1,x_2,...,x_m)={F}_{\textbf{Y}^C}(x_1,x_2,...,x_m)$. So $\rho(\textbf{X})$ and $\rho(\textbf{Y})$ have the same denominators. On the other hand, $\textbf{X}\leq_c\textbf{Y}$ implies that ${{F}_{\textbf{X}}(x_1,x_2,...,x_m)}\leq{{F}_{\textbf{Y}}(x_1,x_2,...,x_m)}$, which concludes the proof.\\
(iv) Obviously, $E[\prod_{i=1}^{m}(-{X_i})]=(-1)^mE[\prod_{i=1}^{m}X_i]$. For the comonontic vector we find
\begin{eqnarray}
                  E[\prod_{i=1}^{m}(-{X_i})^C]\nonumber&=& E[\prod_{i=1}^{m}F_{-X_i}^{-1}(U)], ~U\sim U(0,1)\\\nonumber
                  &=& E[\prod_{i=1}^{m}(-F_{X_i}^{-1}(1-U))]\\\nonumber
                  &=& E[\prod_{i=1}^{m}-F_{X_i}^{-1}(V)],~V\sim U(0,1)\\\nonumber
                  &=& (-1)^mE[\prod_{i=1}^{m}X_i^C].\nonumber
                \end{eqnarray}
Hence,
\begin{eqnarray}
 \nonumber\rho(-{X_1},-{X_2},...,-{X_m})&=& \frac{E[\prod_{i=1}^{m}(-X_i)]-\prod_{i=1}^{m}E[-X_i]}{E[\prod_{i=1}^{m}(-X_i)^C]-\prod_{i=1}^{m}E[X_i]} \\\nonumber
  &=& \frac{(-1)^mE[\prod_{i=1}^{m}X_i]-(-1)^m\prod_{i=1}^{m}E[X_i]}{(-1)^mE[\prod_{i=1}^{m}X_i^C]-(-1)^m\prod_{i=1}^{m}E[X_i]} \\\nonumber
   &=& \rho(X_1,X_2,...,X_m)\nonumber
\end{eqnarray}
which concludes the proof.
\begin{proposition}
 For any random vector $\textbf{X}$ we have $\rho(\textbf{X})\leq 1$. If $\rho(\textbf{X})=1$, then $\textbf{X}=_d{\textbf{X}^C}$.
\end{proposition}
{\bf Proof.} From the fact that
\begin{eqnarray}
 \nonumber F_{\textbf{X}}(y_1,y_2,...,y_m)&\leq&{\min\{{F_{X_1}(y_1),F_{X_2}(y_2),...,F_{X_m}(y_m)}\}}\\\nonumber
   &=& F_{\textbf{X}}^C(y_1,y_2,...,y_m),~\forall~(y_1,y_2,...,y_m)\in\mathbb{R}^m,\nonumber
\end{eqnarray}
it follows that $E[\prod_{i=1}^{m}X_i]\leq{E[\prod_{i=1}^{m}X_i^C]}$, which implies that  $\rho(\textbf{X})\leq 1$.\\
If $\rho(\textbf{X})=1$, then $E[\prod_{i=1}^{m}X_i]={E[\prod_{i=1}^{m}X_i^C]}$ and ${F}_\textbf{X}(x_1,...,x_m)={F}_\textbf{X}^C(x_1,...,x_m)$, so
$\textbf{X}=_d{\textbf{X}^C}$.
\\\indent The reverse implication for $\rho(\textbf{X})=0$ does not hold in general. Indeed, one can easily construct several non-independent random vectors for which $\rho(\textbf{X})=0$.\\
{\bf Example 3.1.}
Let be $Y$ an rv taking the values $0,~\pi/2$ and $\pi$ with probability 1/3 each. Then, it's easy to see that $X_1=\sin Y$ and $X_2=\cos Y$ are uncorrelated (i.e. $\rho(X,Y)=0$). However, they are not independent since $X_1$ and $X_2$ are functionally connected (by the relation $X_1^2+X_2^2=1$).
\\\indent For a continuous counterexample, take $Z\sim U(0,1)$ and $X_1=\sin Z$, $X_2=\cos Z$. Then,
  \begin{equation}
  E[X_1]=E[X_2]=E[X_1X_2]=0\nonumber
  \end{equation}
 so that $X_1$ and $X_2$ are uncorrelated but not independent since the relation $X_1^2+X_2^2=1$ holds.\\
 \indent For the comonotonic random vector the joint tail distribution function has a similar form as the joint distribution function.
 \begin{proposition}
 For any comonotonic random vector $\textbf{X}^C=(X_1^C,X_2^C,...,X_m^C)$, the joint tail distribution function $\bar{F}_{\textbf{X}^C}(\textbf{x}),~\textbf{x}\in\mathbb{R}^m$ has the following expression:
 \begin{equation*}
   \bar{F}_{\textbf{X}^C}(\textbf{x})=\min{\{\bar{F}_1(x_1),\bar{F}_2(x_2),...,\bar{F}_m(x_m)\}},~\textbf{x}\in\mathbb{R}^m.
 \end{equation*}
 \end{proposition}
{\bf Proof.} From (2.3), for any comonotonic random vector $\textbf{X}^C=(X_1^C,...,X_m^C)$, we have
\begin{eqnarray*}
 \bar{F}_{\textbf{X}^C}(\textbf{x})&=& P(X_1^C>x_1,X_2^C>x_2,...,X_m^C>x_m) \\
                                     &=& P(F_1^{-1}(U)>x_1,F_2^{-1}(U)>x_2,...,F_m^{-1}(U)>x_m)  \\
                                       &=& P(U>F_1(x_1),U>F_2(x_2),...,U>F_m(x_m)) \\
                                       &=& P(U>\max{\{F_1(x_1),F_2(x_2),...,F_m(x_m)\}})\\
                                        &=&1-\max{\{F_1(x_1),F_2(x_2),...,F_m(x_m)\}}\\
                                        &=&\min{\{\bar{F}_1(x_1),\bar{F}_2(x_2),...,\bar{F}_m(x_m)\}},~\textbf{x}\in\mathbb{R}^m
\end{eqnarray*}
which concludes the proof.
 \begin{proposition}
  For any non-negative PUOD random vector \textbf{X} we have $\rho(\textbf{X})\geq 0$.
 \end{proposition}
 {\bf Proof.} From the definition of PUOD we know that $\bar{F}^C_{\textbf{X}}(x_1,...,x_m)\geq\bar{F}_{\textbf{X}}(x_1,...,x_m)\geq{\prod_{i=1}^m\bar{F}_{X_i}(x_i)}$, so the integrands in Proposition 3.1 are non-negative, which concludes the proof.
 \section{ The relationships between the new dependence measure and the classical ones}
\setcounter{equation}{0}

  For $m=2$ we can relate $\rho$ to Pearson correlation coefficient $\gamma$, the comonotonicity coefficient $\kappa$ in Koch and Schepper \cite{KD} and $\rho_c$ defined in Dhaene et al. \cite{2014}.
\begin{proposition}
 For any random couple $(X,Y)$ we have $\rho(X,Y)=\gamma(X,Y)$ if and only if the marginal distributions differ only in location and/or scale parameters.
 \end{proposition}
{\bf Proof.} The couples $(X,Y)$ and $(X^C,Y^C)$ have the same marginals, so
\begin{equation}\label{34}
\rho(X,Y)=\frac{E[XY]-E[X]E[Y]}{E[X^CY^C]-E[X]E[Y]}=\frac{Cov(X,Y)}{Cov(X^C,Y^C)}=\frac{\gamma(X,Y)}{\gamma(X^C,Y^C)}\nonumber
\end{equation}
which equals $\gamma(X,Y)$ if and only if $\gamma(X^C,Y^C)=1$. Hence, $Y^C=aX^C+b$ with probability 1 for some constants $a>0$ and $b\in\mathbb{R}$, and thus $F_Y^{-1}(p)=aF_X^{-1}(p)+b$ from which we can conclude that the marginal distributions differ only in location and/or scale parameters.
\begin{proposition} For any random couple $(X,Y)$, we have $\rho(X,Y)=\rho_c(X,Y)$.
\end{proposition}
{\bf Proof.} The proof is obvious.

\begin{proposition}
For any random non-negative random vector $\textbf{X}=(X_1,X_2,...,X_m)$, we have $\rho(\textbf{X})=\kappa(-\textbf{X})$. Especially, when $m=2$, $\rho(\textbf{X})=\kappa(\textbf{X})$.
\end{proposition}
{\bf Proof.}
For any non-negative random vector $\textbf{X}$,
\begin{eqnarray*}
  F_{-\textbf{X}}(\textbf{x}) &=&  P(-X_1\leq x_1,-X_2\leq x_2,...,-X_m\leq x_m)\\
   &=&P(X_1\geq-x_1,X_2\geq-x_2,...,X_m\geq-x_m) \\
   &=& \bar{F}_{\textbf{X}}(-\textbf{x}),
\end{eqnarray*}
so
\begin{eqnarray*}
  \rho(\textbf{X}) &=& \frac{\int_0^{+\infty}\cdots\int_0^{+\infty}\big(\bar{F}_\textbf{X}(\textbf{x})-\bar{F}_\textbf{X}^I(\textbf{x})\big)d\textbf{x}}
{\int_0^{+\infty}\cdots\int_0^{+\infty}\big(\bar{F}_\textbf{X}^C(\textbf{x})-\bar{F}_\textbf{X}^I(\textbf{x})\big)d\textbf{x}}\\
   &=& \frac{\int_0^{+\infty}\cdots\int_0^{+\infty}\big(\bar{F}_\textbf{X}(-\textbf{x})-\bar{F}_\textbf{X}^I(-\textbf{x})\big)d\textbf{x}}
{\int_0^{+\infty}\cdots\int_0^{+\infty}\big(\bar{F}_\textbf{X}^C(-\textbf{x})-\bar{F}_\textbf{X}^I(-\textbf{x})\big)d\textbf{x}} \\
   &=& \frac{\int_0^{+\infty}\cdots\int_0^{+\infty}\big({F}_{-\textbf{X}}(\textbf{x})-{F}_{-\textbf{X}}^I(\textbf{x})\big)d\textbf{x}}
{\int_0^{+\infty}\cdots\int_0^{+\infty}\big({F}_{-\textbf{X}}^C(\textbf{x})-{F}_{-\textbf{X}}^I(\textbf{x})\big)d\textbf{x}} \\
&=&\kappa(-\textbf{X}).
\end{eqnarray*}
\indent When $m=2$, for any non-negative random couple $(X_1,X_2)$,
\begin{equation}
 \bar{F}_{X_1,X_2}(x_1,x_2)=1-F_{X_1}(x_1)-F_{X_2}(x_2)+F_{X_1,X_2}(x_1,x_2).\nonumber
\end{equation}
Therefore,
\begin{eqnarray}
\rho(X_1,X_2)&=&\frac{\int_{0}^{+\infty}\int_{0}^{+\infty}\big(\bar{F}_{X_1,X_2}(x_1,x_2)-\bar{F}_{X_1}(x_1)\bar{F}_{X_2}(x_2)\big)dx_1dx_2}{\int_{0}^{+\infty}\int_{0}^{+\infty}\big(\bar{F}_{X_1,X_2}^C(x_1,x_2)\nonumber
-\bar{F}_{X_1}(x_1)\bar{F}_{X_2}(x_2)\big)dx_1dx_2} \\\nonumber
       &=&\frac{\int_{0}^{+\infty}\int_{0}^{+\infty}\big(F_{X_1,X_2}(x_1,x_2)-F_{X_1}(x_1)F_{X_2}(x_2)\big)dx_1dx_2}{\int_{0}^{+\infty}\int_{0}^{+\infty}\big(F_{X_1,X_2}^C(x_1,x_2)
       -F_{X_1}(x_1)F_{X_2}(x_2)\big)dx_1dx_2}\\ \nonumber
       &=& \kappa(X_1,X_2)\nonumber
\end{eqnarray}
which concludes the proof.\\
\indent {\bf Remark }
 From last formula we can write
\begin{equation}
\rho(X_1,X_2)=\frac{\int_0^1\int_0^1\big(C(u_1,u_2)-u_1u_2\big)dF_{X_1}^{-1}(u_1)dF_{X_2}^{-1}(u_2)}{\int_0^1\int_0^1\big(\min\{u_1,u_2\}-u_1u_2\big)dF_{X_1}^{-1}(u_1)dF_{X_2}^{-1}(u_2)}.\nonumber \\\nonumber
\end{equation}
So $\rho$ depends not only on the copula, but also on the marginals.\\
{\bf Example 4.1.}
In order to illustrate the phenomenon mentioned above, let us consider the $Farlie$-$Gumbel$-$Morgenstern$ copula which is given by
\begin{equation}
C_\alpha(\textbf{u})=u_1u_2[1+\alpha(1-u_1)(1-u_2)],~\alpha\in[-1,1].\nonumber
\end{equation}
It can be shown that $\rho=\frac{1}{3}\alpha$ when both marginals are $U(0,1)$. So the range of $\rho$ for this family is $[-1/3,1/3]$ which obtains the minimum and the maximum. All other marginals will result in a measure smaller than $1/3$. For example, if we insert $Exp(1)$ marginals in the $Farlie$-$Gumbel$-$Morgenstern$ copula, we have $\rho=\frac{1}{4}\alpha$.\\
\indent However, the new dependence measure is different from $\kappa$ in  \cite{KD} and $\rho_c$ in \cite{2014} for $m\geq3$. The following examples can clarify it. \\
{\bf Example 4.2.}
Let's consider the multivariate $Eyraud$-$Gumbel$-$Morgenstern$ function \cite{C} for $n$-dimension
\begin{eqnarray}
 \nonumber H(x_1,x_2,...,x_n)&=&F_1(x_1)F_2(x_2)\cdots F_n(x_n)[1+\Sigma_{1\leq{j_1}<{j_2}\leq n}\alpha_{j_1j_2}\big(1-F_{j_1}(x_{j_1})\big)\big(1-F_{j_2}(x_{j_2})\big)\\\nonumber
                        &&+\cdots+ \alpha_{12...n}\big(1-F_1(x_1)\big)\big(1-F_2(x_2)\big)\cdots \big(1-F_n(x_n)\big)]\nonumber
\end{eqnarray}
 where the coefficients $\alpha's$ are real constants. For the marginals are $U(0,1)$, after complicated calculation we can get the following results for $n=3$
\begin{eqnarray}
   \nonumber \rho(U_1,U_2,U_3)&=&\frac{E[U_1U_2U_3]-E[U_1]E[U_2]E[U_3]}{E[U_1^CU_2^CU_3^C]-E[U_1]E[U_2]E[U_3]}\\\nonumber
   \nonumber &=&\frac{\frac{1}{8}+\frac{1}{72}(\alpha_{12}+\alpha_{23}+\alpha_{13})-\frac{1}{216}\alpha_{123}-\frac{1}{8}}{\frac{1}{4}-\frac{1}{8}}\\
    &=&\frac{1}{9}(\alpha_{12}+\alpha_{23}+\alpha_{13})-\frac{1}{27}\alpha_{123}.\nonumber
      \end{eqnarray}
      However,
\begin{eqnarray}
       \nonumber\kappa(U_1,U_2,U_3)&=&\frac{\int_{0}^{1}\int_{0}^{1}\int_{0}^{1}\big(F_{\textbf{U}}(x_1,x_2,x_3)-F_{U_1}(x_1)F_{U_2}(x_2)F_{U_3}(x_3)\big)dx_1dx_2dx_3}{\int_{0}^{1}\int_{0}^{1}\int_{0}^{1}\big(\min\{{F_{U_1}(x_1),F_{U_2}(x_2),F_{U_3}(x_3)}\}-F_{U_1}(x_1)F_{U_2}(x_2)F_{U_3}(x_3)\big)dx_1dx_2dx_3}\nonumber \nonumber\\&=&\frac{\frac{1}{72}(\alpha_{12}+\alpha_{23}+\alpha_{13})+\frac{1}{216}\alpha_{123}}{\frac{1}{8}}\nonumber \\
        &=&\frac{1}{9}(\alpha_{12}+\alpha_{23}+\alpha_{13})+\frac{1}{27}\alpha_{123}\nonumber.
\end{eqnarray}
{\bf Example 4.3.}
Suppose non-negative random vector $\textbf{X}$ obeys a multivariate Pareto distribution of the second kind introduced by Alexandru et al. \cite{2010}. If $\textbf{X}\thicksim Par_m(II)(\boldsymbol{\mu},\boldsymbol{\sigma},\boldsymbol{\alpha},\alpha_0)$, then the decumulative distribution function is
\begin{equation*}
  \bar{F}_{\textbf{X}}(\textbf{x})=(1+\max_{j=1}^m\frac{x_j-\mu_j}{\sigma_j})^{-\alpha_0}\prod_{j=1}^m(1+\frac{x_j-\mu_j}{\sigma_j})^{-\alpha_j},~
  x_j>\mu_j,~j=1,2,...,m,
\end{equation*}
where $\mu_j$ are real and $\sigma_j$, $\alpha_j$, $\alpha_0$ are positive constants. The marginal distribution is
\begin{equation*}
  \bar{F}_{X_j}(x)=(1+\frac{x_j-\mu_j}{\sigma_j})^{-\alpha_j},~ x_j>\mu_j, ~\sigma_j>0, ~\alpha_j>0,
\end{equation*}
which is a heavy tail distribution. In addition, setting $\alpha_0=0$ yields a probabilistic model having independent Pareto distributed marginals. In order to simplify our calculation, we set $\textbf{X}\thicksim Par_m(II)(\textbf{0},\textbf{1},\boldsymbol {\alpha},\alpha_0)$, where $\boldsymbol {\alpha}=(\alpha,...,\alpha)$, i.e. the marginal distributions are same.\\
\indent After complicated calculation, we can get the results when $m=3,~\alpha>3$ as follows.
\begin{equation*}
  E[X_j]=\frac{1}{\alpha_0+\alpha-1},~j=1,2,3,
\end{equation*}
\begin{equation*}
 E[X_1X_2X_3]=\frac{6}{(\alpha_0+\alpha-1)(\alpha_0+2\alpha-2)(\alpha_0+3\alpha-3)},
\end{equation*}
\begin{equation*}
  E[X_1^CX_2^CX_3^C]=\frac{1}{(\alpha-1)(\alpha-2)(\alpha-3)}.
\end{equation*}
Therefore, we have
\begin{eqnarray*}
  \rho(X_1,X_2,X_3)&=&\frac{E[X_1X_2X_3]-E[X_1]E[X_2]E[X_3]}{E[X_1^CX_2^CX_3^C]-E[X_1]E[X_2]E[X_3]}\\
  &=&\frac{\frac{6}{(\alpha_0+\alpha-1)(\alpha_0+2\alpha-2)(\alpha_0+3\alpha-3)}-{\frac{1}
  {(\alpha_0+\alpha-1)^3}}}{\frac{1}{(\alpha-1)(\alpha-2)(\alpha-3)}-{\frac{1}{(\alpha_0+\alpha-1)^3}}}.
 \end{eqnarray*}
 \indent For the covariances, we get the following results.
 \begin{equation*}
   \text{Cov}(X_i,X_j)=\frac{\alpha_0}{(\alpha_0+\alpha-1)^2(\alpha_0+2\alpha-2)},~i,j=1,2,3,
 \end{equation*}
 \begin{equation*}
  \text{Cov}(X_i^C,X_j^C)=\frac{\alpha_0+\alpha}{(\alpha_0+\alpha-1)^2(\alpha_0+\alpha-2)},~i,j=1,2,3.
 \end{equation*}
 So \begin{eqnarray*}
      \rho_C(X_1,X_2,X_3) &=& \frac{\sum_{i=1}^3\sum_{j<i}\text{Cov}(X_i,X_j)}{\sum_{i=1}^3\sum_{j<i}\text{Cov}(X_i^C,X_j^C)} \\
       &=& \frac{\alpha_0(\alpha_0+\alpha-2)}{(\alpha_0+\alpha)(\alpha_0+2\alpha-2)}.
    \end{eqnarray*}
\begin{figure}[htb]
    \centering
     \includegraphics[scale=1]{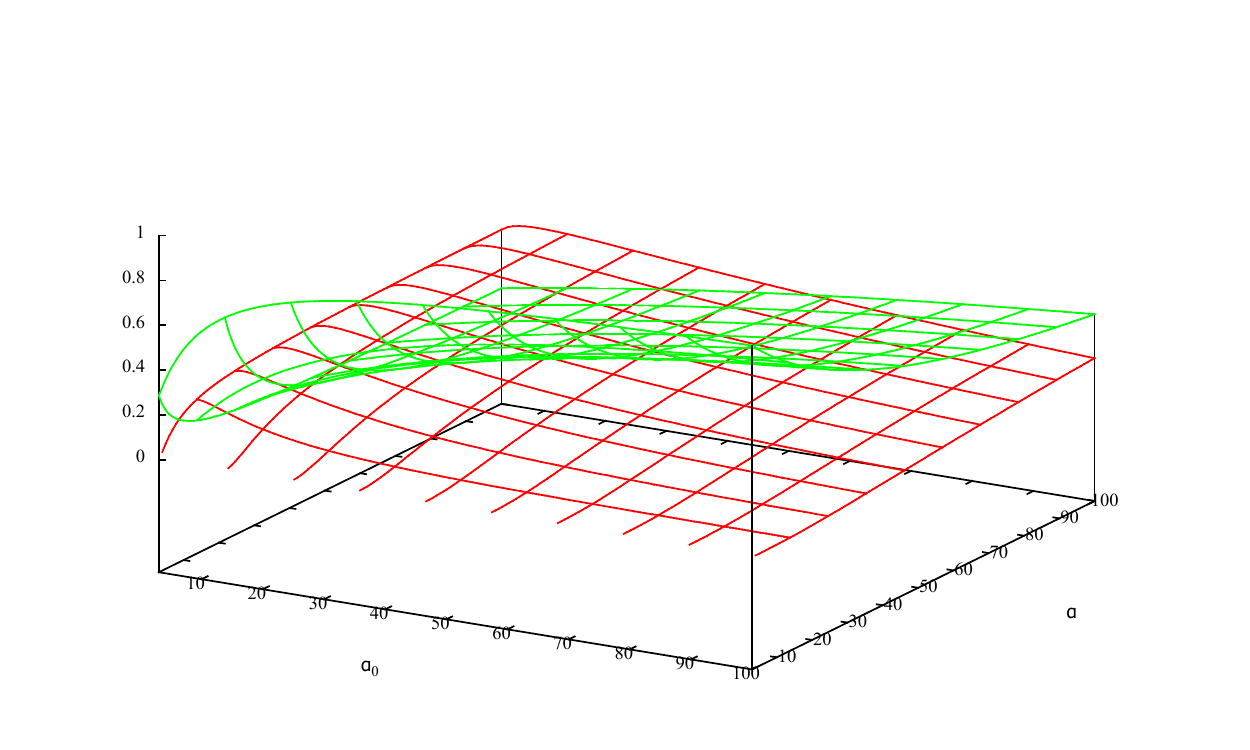}
    \caption{$\rho(X_1,X_2,X_3)$ (the red) and $\rho_C(X_1,X_2,X_3)$ (the green) for $(X_1,X_2,X_3)\thicksim Par_3(II)(\textbf{0},\textbf{1},(\alpha,\alpha,\alpha),\alpha_0),~\alpha>3$.}
\end{figure}
\indent We can give the figure of $\rho$ and $\rho_C$ in Figure 1. From the figure, it's not difficult to know that this multivariate Pareto distribution can't be used to describe the comonotonic case no matter what $\alpha_0$ and $\alpha$ are.  \\
{\bf Example 4.4.}
Suppose that risks
$$\textbf{X}=(X_1,X_2,...,X_m)'\thicksim Nor((\mu_1,\mu_2,...,\mu_m)',\Sigma),$$
$$~\Sigma=\left(
 \begin{array}{ccc}
  \sigma_1^2 & \sigma_{12} ~~\cdots &\sigma_{13} \\
   \sigma_{12} &\sigma_2^2  ~~~\cdots &\sigma_{23} \\
   \vdots    &\vdots   \   \   \  &\vdots    \\
   \sigma_{1m} &\sigma_{2m} ~~\cdots &\sigma_m^2 \\
   \end{array}
   \right),
   $$
   then
  $$\textbf{X}^C=(X_1^C,X_2^C,...,X_m^C)'\thicksim Nor((\mu_1,\mu_2,...,\mu_m)',\Sigma^C),$$
  $$\Sigma^C=\left(
  \begin{array}{ccc}
    \sigma_1^2 & \sigma_1\sigma_2 ~~\cdots &\sigma_1\sigma_m \\
    \sigma_1\sigma_2  &\sigma_2^2 ~~\cdots  &\sigma_2\sigma_m \\
     \vdots    &\vdots   \   \   \  &\vdots    \\
    \sigma_1\sigma_m & \sigma_2\sigma_m ~~\cdots &\sigma_m^2  \\
    \end{array}
     \right),$$
  where $\sigma_{ij}=\rho_{ij}\sigma_i\sigma_j,~i,j=1,2,...,m$.\\
\indent In order to calculate the product moments easily, we can refer to the characteristic function
\begin{equation}
    \varphi_{\textbf{X}}(\textbf{t})=e^{i{\textbf{t}'}\boldsymbol{\mu}-\frac{1}{2}\textbf{t}'\Sigma\textbf{t}}\nonumber\\
\end{equation}
for $\textbf{X}\thicksim Nor(\boldsymbol{\mu},\Sigma)$ and $\textbf{t}\in\mathbb{R}^m$. \\
 \indent As is well known, for any random vector $\textbf{X}$, the characteristic function $\varphi_{\textbf{X}}(\textbf{t})$ always exists. So suppose that for a random vector $\textbf{X}$ the expectation $E[\prod_{i=1}^{m}X_i^{r_k}]$ exists for some set of non-negative integers $r_1,r_2,...,r_m$. Then this expectation can be found from the relation
\begin{equation*}
 E[\prod_{i=1}^{m}X_i^{r_k}]=\frac{1}{i^{r_1+r_2+\cdots+r_m}}\left(\frac{\partial^{r_1+r_2+\cdots+r_m}}
 {\partial^{r_1}t_1\partial^{r_2}t_2\cdots\partial^{r_m}t_m}\varphi_{\textbf{X}}(\textbf{t})\right)|_{\textbf{t}=\textbf{0}}
\end{equation*}
where $\textbf{0}=(0,0,...,0)'$.\\
\indent  In this example we can get the following characteristic functions for $\textbf{X}$ and $\textbf{X}^C$
\begin{eqnarray*}
   \varphi_{\textbf{X}}(\textbf{t}) &=& e^{i\sum_{i=1}^m\mu_it_i-\frac{1}{2}\sum_{i=1}^m\sum_{j=1}^m\rho_{ij}\sigma_{i}\sigma_{j}t_it_j}, \\
   \varphi_{\textbf{X}^C}(\textbf{t}) &=& e^{i\sum_{i=1}^m\mu_it_i-\frac{1}{2}\sum_{i=1}^m\sum_{j=1}^m\sigma_{i}\sigma_{j}t_it_j}.
 \end{eqnarray*}
 Differentiate the characteristic function with respect to $\textbf{t}$ and let $\textbf{t}=\textbf{0}$, we can get the following product moments:
 \begin{eqnarray*}
 E[\prod_{i=1}^mX_i] &=& (-i)^m\frac{\partial^m\varphi_{\textbf{X}}(\textbf{t})}{\partial t_1\partial t_2...\partial t_m}|_{\textbf{t}=\textbf{0}}\\
                     &=&\prod_{k=1}^m\mu_{j_k}+\sum_{S(m,1)}\sigma_{j_1j_2}\prod_{k=3}^m\mu_{j_k}+\sum_{S(m,2)}\sigma_{j_1j_2}\sigma_{j_3j_4}\prod_{k=5}^m\mu_{j_k}+\cdots\\
                     &&+\sum_{S(m,\frac{2m-5+(-1)^m}{4})}\sigma_{j_1j_2}\cdots\sigma_{j_Aj_B}\prod_{k=C}^m\mu_{j_k}
                     +\sum_{S(m,\frac{2m-3+(-1)^m}{4})}\sigma_{j_1j_2}\cdots\sigma_{j_Dj_E}\prod_{k=F}^m\mu_{j_k}
\end{eqnarray*}
where $A=m-4+\frac{1+(-1)^m}{2},~B=m-3+\frac{1+(-1)^m}{2},~C=D=m-2+\frac{1+(-1)^m}{2},~E=m-1+\frac{1+(-1)^m}{2},~F=\frac{[m-(-1)^mm]}{2},
~S(m,k)=\frac{m!}{2^kk!(m-k)!}$ and $\sum_{S(m,k)}$ means the sum of S(m,k) cases; see \cite{20}.\\
\indent Therefore,
 \begin{equation*}
   \rho_C(\textbf{X})=\frac{\sum_{i=1}^m\sum_{j<i}\text{Cov}(X_i,X_j)}{\sum_{i=1}^m\sum_{j<i}\text{Cov}(X_i^C,X_j^C)}
   =\frac{\sum_{i=1}^m\sum_{j<i}\rho_{ij}\sigma_i\sigma_j}{\sum_{i=1}^m\sum_{j<i}\sigma_i\sigma_j},
\end{equation*}
\begin{equation*}
\rho(\textbf{X})=\frac{E[\prod_{i=1}^mX_i]-\prod_{i=1}^mE[X_i]}{E[\prod_{i=1}^mX^C_i]-\prod_{i=1}^mE[X_i]} =\frac{I}{J}
\end{equation*}
where
\begin{eqnarray*}
 I&=&\sum\limits_{S(m,1)}\rho_{j_1j_2}\sigma_{j_1}\sigma_{j_2}\prod\limits_{k=3}\limits^m\mu_{j_k}+\sum\limits_{S(m,2)}\rho_{j_1j_2}\sigma_{j_1}\sigma_{j_2}\rho_{j_3j_4}\sigma_{j_3}\sigma_{j_4}\prod\limits_{k=5}\limits^m\mu_{j_k}+\cdots
                     \\&&+\sum\limits_{S(m,\frac{2m-5+(-1)^m}{4})}\rho_{j_1j_2}\sigma_{j_1}\sigma_{j_2}\cdots\rho_{j_Aj_B}\sigma_{j_A}\sigma_{j_B}\prod\limits_{k=C}\limits^m\mu_{j_k}
                     \\&&+\sum\limits_{S(m,\frac{2m-3+(-1)^m}{4})}\rho_{j_1j_2}\sigma_{j_1}\sigma_{j_2}\cdots\rho_{j_Dj_E}\sigma_{j_D}\sigma_{j_E}\prod\limits_{k=F}\limits^m\mu_{j_k},
  \end{eqnarray*}
\begin{eqnarray*}
J&=&\sum\limits_{S(m,1)}\sigma_{j_1}\sigma_{j_2}\prod\limits_{k=3}\limits^m\mu_{j_k}+\sum\limits_{S(m,2)}\sigma_{j_1}\sigma_{j_2}\sigma_{j_3}\sigma_{j_4}\prod\limits_{k=5}\limits^m\mu_{j_k}+\cdots
                 \\&& +\sum\limits_{S(m,\frac{2m-5+(-1)^m}{4})}\sigma_{j_1}\sigma_{j_2}\cdots\sigma_{j_Aj_B}\prod\limits_{k=C}\limits^m\mu_{j_k}
                 +\sum\limits_{S(m,\frac{2m-3+(-1)^m}{4})}\sigma_{j_1}\sigma_{j_2}\cdots\sigma_{j_Dj_E}\prod\limits_{k=F}\limits^m\mu_{j_k},
  \end{eqnarray*}
 and  $A=m-4+\frac{1+(-1)^m}{2},~B=m-3+\frac{1+(-1)^m}{2},~C=D=m-2+\frac{1+(-1)^m}{2},~E=m-1+\frac{1+(-1)^m}{2},~F=\frac{[m-(-1)^mm]}{2},
~S(m,k)=\frac{m!}{2^kk!(m-k)!}$.\\
\indent {\bf Remark 4.1.} It's obvious that $\rho(\textbf{X})$ not only depends on $\sigma_{ij}$ but also on $\mu_j$ while $\rho_C(\textbf{X})$ only relys on $\sigma_{ij}$, $i,j=1,2,...,m$. Especially,
when $m=3$ and $\mu_1=\mu_2=\mu_3$,
\begin{eqnarray*}
\rho(\textbf{X})=\frac{\rho_{12}\sigma_1\sigma_2+\rho_{13}\sigma_1\sigma_3+\rho_{23}\sigma_2\sigma_3}{\sigma_1\sigma_2+\sigma_1\sigma_3+\sigma_2\sigma_3}
=\rho_C(\textbf{X}).
\end{eqnarray*}

\section{Estimation}
\setcounter{equation}{0}

 \indent In this section we give some comments on the estimation of $\rho$ from a sample of $\textbf{X}$. Inference on $\rho(\textbf{X})$ in Definition 3.1 boils down to inference on the expectations $E[\prod_{i=1}^{m}X_i]$, $E[\prod_{i=1}^{m}X_i^C]$ and $\prod_{i=1}^{m}E[X_i]$. These can be estimated by moments estimation. This is of great significance in a real-life data set when the product moment is difficult to find. \\
\indent Suppose we want to estimate the new dependence measure for $m$ variables, for which we have $n$ coupled observations, not necessarily independent, summarized in a data matrix Y (dimension $n\times m$).\\
 \indent So we can get the the estimations
 \begin{equation*}
\widehat{E}[\prod_{j=1}^{m}X_j]= \frac{1}{n}\sum_{i=1}^n\prod_{j=1}^mY_{ij},
 \end{equation*}
 \begin{equation*}
 \widehat{E}[X_j]= \frac{1}{n}\sum_{i=1}^nY_{ij}.
 \end{equation*}
\indent For the estimation of $E[\prod_{j=1}^{m}X_j^C]$, we need a sample of $\textbf{X}^C$. Dhaene et al.[8] show that for any $\textbf{x}$ and $\textbf{y}$ in the range of a comonotonic vector either $\textbf{x}\leq\textbf{y}$ or $\textbf{x}\geq\textbf{y}$ holds. In other words, all possible outcomes of $\textbf{X}^C$ are ordered componentwise. As $\textbf{X}$ and $\textbf{X}^C$ also have the same marginal distributions, we can easily turn the sample of $\textbf{X}$ into a sample of $\textbf{X}^C$. Denoting the $i$th order statistic of $X_j$ by $Y_j^{(i)}$, we find the following sample of $\textbf{X}^C$: ${(Y_1^{(i)},...,Y_m^{(i)})},~i=1,2,...,n$. Accordingly, we can get the product moment estimation
\begin{equation*}
 \widehat{E}[\prod_{j=1}^{m}X_j^C]=\frac{1}{n}\sum_{i=1}^n\prod_{j=1}^mY_j^{(i)}.
\end{equation*}
Summarizing, we find the following estimator for $\rho$ :
\begin{equation*}
\widehat{\rho}(\textbf{X})=\frac{\frac{1}{n}\sum_{i=1}^n\prod_{j=1}^mY_{ij}-\prod_{j=1}^m\frac{1}{n}\sum_{i=1}^nY_{ij}}
{\frac{1}{n}\sum_{i=1}^n\prod_{j=1}^mY_j^{(i)}-\prod_{j=1}^m\frac{1}{n}\sum_{i=1}^nY_{ij}}.
\end{equation*}
\indent When the random vectors are non-negative, the estimation has another form. Actually, the empirical tail distribution can be written as
 \begin{equation*}
   \hat{\bar{F}}_\textbf{X}(x_1,x_2,...,x_m)=P(X_1>x_1,X_2>x_2,...,X_m>x_m)=\frac{1}{n}\sum_{i=1}^n\prod_{j=1}^mI(Y_{ij}>x_j);
 \end{equation*}
 for the empirical versions of the distribution for independent and comonotonic vectors of the same Fr$\acute{e}$chet space,we have
 \begin{equation*}
  \hat{\bar{F}}_\textbf{X}^I(x_1,x_2,...,x_m)=\prod_{j=1}^m\hat{\bar{F}}_j(x_j)=\prod_{j=1}^m(\frac{1}{n}\sum_{i=1}^nI(Y_{ij}>x_j)),
 \end{equation*}
 \begin{equation*}
   \hat{\bar{F}}_\textbf{X}^C(x_1,x_2,...,x_m)=P(X_1^C>x_1,X_2^C>x_2,...,X_m^C>x_m)=\frac{1}{n}\sum_{i=1}^n\prod_{j=1}^mI(Y_{j}^{(i)}>x_j).  \\
 \end{equation*}
Now, define $m_j= \min_{i=1}^n Y_{ij}$ and $M_j= \max_{i=1}^n Y_{ij}$. As a consequence, $Y_j^{(1)}= m_j$ and $Y_j^{(n)}= M_j$ for any $j=1,2,...,m,$ and $\int_{m_j}^{M_j}I(Y_{ij}>x_j)dx_j=Y_{ij}-m_j$ for any $i=1,2,...,n,~j=1,2,...,m.$\\
 \indent If we replace the tail distribution functions in Proposition 3.1 by their empirical versions, we get the estimator for $\rho$ :
 \begin{equation*}
\widehat{\rho}(\textbf{X})=\frac{\frac{1}{n}\sum_{i=1}^n\prod_{j=1}^m(Y_{ij}-m_j)-\frac{1}{n^m}\prod_{j=1}^m(\sum_{i=1}^nY_{ij}-nm_j)}
{\frac{1}{n}\sum_{i=1}^n\prod_{j=1}^m(Y_j^{(i)}-m_j)-\frac{1}{n^m}\prod_{j=1}^m(\sum_{i=1}^nY_{ij}-nm_j)}.
\end{equation*}
%
%

\noindent{\bf Acknowledgements.} \ 
The research   was supported by the National
Natural Science Foundation of China (No. 11171179),  the Research
Fund for the Doctoral Program of Higher Education of China (No. 20133705110002) and the Program for  Scientific Research Innovation Team in Colleges and Universities of Shandong Province.

\end{document}